\documentclass[article,twocolumn]{aastex61}
\pdfoutput=1 
\usepackage{amsmath,amstext}
\usepackage{apjfonts}
\usepackage{afterpage}

\usepackage{xcolor}
\include{bibdef}
\usepackage{lineno}

\def\et3{\eta_3}
\def\th1{\theta_{-1}}
\def\r07{r_{0,7}}
\def\x05{x_{0.5}}

\def\cm{\hbox{~cm}}

\def\Mpc{\hbox{~Mpc}}

\def\s{\hbox{~s}}

\def\Hz{\hbox{~Hz}}

\def\erg{\hbox{~erg}}
\def\s{{\hbox{~s}}
\def\cm2{\hbox{~cm}^2}}
\usepackage{xspace}

\def\gwgrb{GW150914-GBM\xspace}
\def\gw{GW150914\xspace}
\def\bbgr{BBH-to-GRB ratio\xspace}

\begin{document}


\title{Fermi GBM follow-up of LIGO-Virgo Binary black hole mergers - detection prospects}
\author[0000-0002-2149-9846]{P. Veres}
\affiliation{Center for Space Plasma and Aeronomic Research, University of
Alabama in Huntsville, 320 Sparkman Drive, Huntsville, AL 35899, USA}
\author[0000-0001-5078-9044]{T.~\surname{Dal Canton}}
\altaffiliation{NASA Postdoctoral Program fellow}
\affiliation{NASA Goddard Space Flight Center, Greenbelt, MD 20771, USA}
\author{E.~Burns}
\altaffiliation{NASA Postdoctoral Program fellow}
\affiliation{NASA Goddard Space Flight Center, Greenbelt, MD 20771, USA}
\author{A.~Goldstein}
\affiliation{Science and Technology Institute, Universities Space Research Association, Huntsville, AL 35805, USA}

\author{T.~B.~Littenberg}
\affiliation{NASA Marshall Space Flight Center, Huntsville, AL 35811, USA}

\author{N.~Christensen}
\affiliation{Carleton College, Northfield, MN 55057, USA}
\affiliation{Artemis, Universit\'e C\^ote d'Azur, Observatoire C\^ote d'Azur, CNRS, CS 34229, F-06304 Nice Cedex 4, France}

\author[0000-0003-1626-7335]{R.~D.~Preece}
\affiliation{Center for Space Plasma and Aeronomic Research, University of Alabama in Huntsville, 320 Sparkman Drive, Huntsville, AL 35899, USA}

\begin{abstract}
{\it Fermi}-Gamma-ray Burst Monitor  observed a 1 s long gamma-ray signal
(\gwgrb) starting 0.4 s after the first gravitational wave detection from the
binary black hole  merger GW150914. \gwgrb is consistent with a short gamma-ray
burst origin; however, no unambiguous claims can be made as to the physical association
of the two signals due to a combination of low gamma-ray flux and unfavorable
location for {\it Fermi}-GBM.  Here we answer the following question: if GW150914 and
GW150914-GBM were associated, how many LIGO-Virgo  binary black hole mergers
would {\it Fermi}-GBM have to follow up to detect a second source? To answer
this question, we perform simulated observations of binary black hole mergers with LIGO-Virgo
and adopt different scenarios for gamma-ray emission from the literature. We
calculate the ratio of simulated binary black hole mergers detected by LIGO-Virgo to the
number of gamma-ray counterpart detections by {\it Fermi}-GBM, \bbgr.  A large
majority of the models considered here predict a \bbgr in the range of 5 to 20,
but for optimistic cases can be as low as 2 or for pessimistic assumptions as
high as 700.   Hence we expect that the third observing run, with its high rate of binary black hole detections and assuming the absence of a joint detection, will provide strong constraints on the presented models.
\end{abstract}

\section{Introduction} \label{sec:intro}

Gravitational waves (GWs) were directly observed for the first time (GW150914)
from the coalescence of two black holes \citep[BH; ][]{Abbott+16gw150914},
opening a new window on the universe.  There is an intense, ongoing effort to
detect electromagnetic (EM) counterparts for GW observations
\citep{Abbott+16followup, Racusin+17gwfup, Abbott+17joint170817a, 2019ApJ...871...90B,Abbott+19followup}. This effort resulted in the
detection of the first unambiguous gamma-ray  counterpart to a GW 
signal \citep{Goldstein+17170817,GBMLVC17, Abbott+17joint170817a}, from a
binary neutron star merger.

{\it Fermi} Gamma-ray Burst Monitor (GBM) observations around GW150914 uncovered a 1 s long gamma-ray
signal (designated \gwgrb),  0.4 s after GW150914, having a broadly consistent sky
location \citet{Connaughton+16gbmgw,Connaughton+18gw2}. This event did not
trigger GBM, but was found in an off-line search
\citep[e.g.][]{Blackburn+15targeted, Goldstein+16updates, Kocevski+18targeted,
Goldstein+19O3update}.  Having been identified using a hard spectral template
and lasting $\sim$ 1 s, \gwgrb is consistent with a short GRB.  The chance
association between GW150914 and GW150914-GBM (P=$2.2\times 10^{-3}$,
\citet{Connaughton+16gbmgw}) is not low enough to claim common origin for the
two signals. {\it Fermi}-GBM routinely follows up BBH merger events. So far there has been no unambiguous gamma-ray counterpart detection \citep[e.g.][]{Goldstein+17gcn, Burns+15gcn, Hamburg+17gcn}.

An EM counterpart to a stellar mass binary BH (BBH) merger is not
commonly expected, however there are a wealth of proposed mechanisms that could
at least in principle provide detectable emission. Further observations in GWs
and gamma-rays are needed to confirm or render such an association unlikely. In this
paper we calculate the ratio of  BBH mergers to gamma-ray counterparts for
{\it Fermi}-GBM in different scenarios, assuming the GW150914 association is real. We
calculate concrete estimates for future BBH observations during the ongoing
LIGO-Virgo third observing run (O3).

A large number of scenarios have been outlined for producing observable
gamma-rays from BBH mergers.  \citet{Loeb16gw} sketches a scenario where a
rapidly rotating core of a massive star collapses into two BHs that merge and
accrete the remaining stellar material and produce \gwgrb. \citet{Dai+17loebgw}
argues that the heat associated with dynamical friction for the BBH would
remove the star as a potential accretion source, unless the BBH forms close to
the center. In this case the dynamical friction speeds up the merger which has
observable imprints on the waveform \citep{Fedrow+17loebgw}.
\citet{Perna+16gw} propose a disk around one of the BHs that survives up to the
time of the merger and it is subsequently accreted. This scenario was also
discussed by \citet{Woosley16bbh} and challenged by \citet{Kimura+17disk}.
\citet{Zhang16chargedBH} presents a scenario, where charged BHs orbit each
other before merging, \citet{Lyutikov16gw} argues however that the required
charge is prohibitive. BBHs associated with GRBs can also be used to constrain
GRB emission mechanisms \citep{Veres+16gwgrb}.

In considering the possible reasons why other observed BBH mergers did not
produce gamma-rays, aside from the uncertainty on the physical mechanism, we
note that \gw is one of the most massive BBH merger observed. Together with the
unfavorable position for {\it Fermi}-GBM, \gwgrb was observed close to the
sensitivity limit.  If there is a positive correlation between the total mass and
the EM output, it is not surprising that there are no EM counterparts observed
for lower mass or more distant BBH mergers.

Here we attempt to answer the following question: if \gw and \gwgrb  were
indeed associated, 
how many LIGO-Virgo BBH mergers will {\it Fermi}-GBM
 have to follow-up to detect a second source.  We perform simulated observations of BBH mergers
with LIGO-Virgo, adopt different scenarios for gamma-ray emission and calculate
the ratio of BBH mergers detected by LIGO-Virgo to the number of gamma-ray
counterpart detections by {\it Fermi}-GBM.  An alternative formulation of the above
question is: given an emission model, 
how many BBH follow-up observations (that are
 non-detections) are needed
 to rule out the specific model for \gwgrb.

The exposure function of GBM depends on a number of factors.  Any given point
on the sky will be observed roughly 60-70\% of the time. Here we take 65\% as
an average. The two main reasons of the loss of coverage are: the source being
occulted by Earth and the fact that Fermi is turned off during passage through
the South Atlantic Anomaly.

We structure the paper in the following way: in Section \ref{sec:gw} we
describe the simulation of BBHs, in Section \ref{sec:em} we describe possible
scenarios by which gamma-rays and GWs can be related. We discuss our results in
Section \ref{sec:disc} and conclude in Section \ref{sec:concl}. Physical
constants have their usual meaning.  For cosmological calculations we use
$\Omega_m=1-\Omega_\Lambda=0.286$ and $H_0= 69.6~ {\rm km} \s^{-1} \Mpc^{-1}$
\citep{Bennett+14cosmo}.

\section{Binary black hole merger simulations} \label{sec:gw} 

\subsection{Mass and spin assumptions}
\label{sec:mass}
We simulate the primary mass of the BBH systems ($M_1$) following a power law
distribution \citep{Kovetz+17massfunction} $P(M_1)\propto M_1^{\alpha_M}$
between $M_{\rm min}=5 M_{\odot}$ and $M_{\rm max}= 100 M_{\odot}$ and
$\alpha_M=-2.35 $ from estimates in \citet{Abbott+17gw170104}. At high masses
the distribution is cut by an exponential function, so  the differential
distribution is $dN/dM_1=C \left(M_1/5 M_\odot\right)^{\alpha_M}
\exp{(-M_1/M_{\rm cut})}$ where C is a constant, and $M_{\rm cut}=60
M_{\odot}$. The secondary mass ($M_2$) follows a uniform distribution between
$M_{\rm min}$ and M$_1$.  

The spins of individual BHs in the first Gravitaitonal Wave Transient Catalog \citep[GWTC-1, ][]{GWTC1} are not extreme, and thus they will not have a dominant effect in determining the spin parameter of the final BH. Based on this observation, we use non-spinning BH components in our simulations.
In addition, the distribution of final spins in from our simulations is consistent with the final spins presented in GWTC-1.

\subsection{Spatial distribution} \label{sec:spatial}
We generate a population of binary black hole systems and analyze their
LIGO-Virgo signal.  The spatial distribution of BBHs is uniform in volume up to a
distance D$_{\rm max}$. D$_{\rm max}$ is determined from the requirement that
all systems up to this distance with optimal viewing properties (face-on or
inclination angle, $\iota=0$) should be detectable at signal to noise ratio (SNR, see Section \ref{sec:snr}) levels corresponding
to at least borderline detection (e.g.  SNR$>$8). This guarantees that we account
for every BBH system potentially detectable by LIGO-Virgo for a given
sensitivity.  Within 4 Gpc, a BBH at the high end of the mass distribution with a
favorable inclination, will be detected marginally. We thus fix   D$_{\rm max}=$ 4
Gpc.

We assign a total angular momentum vector to every BBH pointing at random
directions.  This will correspond to the direction the jet is launched, and
also defines the inclination angle, $\iota$ as seen from Earth. Any non-zero individual spins would have an effect on the direction total angular momentum. 

\subsection{SNR calculation}
\label{sec:snr}
We use the {\tt LALsim-inspiral}
tool\footnote{\url{https://www.lsc-group.phys.uwm.edu/daswg/docs/howto/lal-install.html}}
and generate frequency domain waveforms for the simulated BBH mergers
\citep{Jimenez+17gwfit,Husa+16waveforms,Khan+16gwform}. BBHs are uniformly distributed in a
$D_{\rm max}=$4 Gpc radius sphere. The masses that enter LALsim-inspiral are
$M_{\rm det}=M_{\rm source} (1+z)$, where $M_{\rm source}$ are the simulated
masses and $z$ is the redshift. We take the
generated $h_+$ and $h_\times$ waveforms, and using equations 17-18 of
\citet{Schutz11FoM}, we convolve them with the antenna responses $F_+$ and
$F_\times$ for Hanford and Livingston to obtain the simulated SNR:

\begin{equation}\label{eq:snr}
{\rm SNR}^2 =4 \int \frac{|h_+ F_+ + h_\times F_\times|^2}{S_h(f)} df.
\end{equation}

We integrate the waveform from $f_{\rm low}=20\Hz$  and take the lower of the
two instruments' SNR. If it is above 8, we consider it a detection.
Alternatively we consider SNR of 12 as a more conservative detection threshold.
Past and future estimated sensitivity ($S_h(f)$) of LIGO are compiled in
\citep{Abbott+16GWsens}, we use the 'Mid high/Late low' sensitivity for both
Hanford and Livingston as a proxy for O3 sensitivity.

We need a large sample of high SNR GW detections to have meaningful statistics.
E.g. for $\approx1000$ BBHs with SNR $>$ 8 with the  mass distribution given in
Section \ref{sec:mass} out to D$_{\rm max}$= 4 Gpc, we need $\sim$400,000 simulated
BBHs. Because most of these systems will not contribute above SNR = 8, in order
to speed up the calculation, we do a preliminary SNR calculation (SNR$_{\rm
S}$) using average antenna functions and a simplified waveform.  For a BBH with
total mass $M_t$, and inclination angle $\iota$ to our line of sight, the signal
to noise ratio measured by LIGO-Virgo can be approximated as \citep{Dalal+06snr,OLeary+16snr}:
\begin{equation}
{\rm SNR^2_{\rm S}} = 4 \frac{\mathcal{A}^2}{D_L^2}
\left(\left<F_+^2\right>(1+\cos^2\iota)^2+4\left<F_\times^2\right> \cos^2
\iota\right) \int\limits_{f_{\rm low}}^{f_{\rm ISCO}} \frac{f^{-7/3}}{S_h(f)}
df,
\end{equation}
where $\mathcal{A}=\sqrt{\frac{5}{96}} \pi^{-2/3} \left(\frac{G
M_t}{c^3}\right)^{5/6}c $, $D_L$ is the luminosity distance, and $F_+$ and
$F_\times$ are taken to be their square-averaged value 1/5. 
$f_{\rm ISCO}(M_t)= c^3 6^{-3/2} \pi^{-1} G^{-1} M_t^{-1}\approx$ $ 438
(M/10M_\odot)^{-1} \Hz$ is the GW frequency at the innermost
stable circular orbit around the final BH. 

First we calculate the simple waveform and if it indicates an SNR$_{\rm S}>$5,
we subsequently calculate the SNR using the advanced waveform. By this method
we only need to calculate 5\% of the total cases using the advanced waveform.
The chance of having an SNR$_{\rm S}<5$ but SNR$>8$ is less than $10^{-3}$.

\begin{figure}
\centering
\includegraphics[width=0.9\columnwidth]{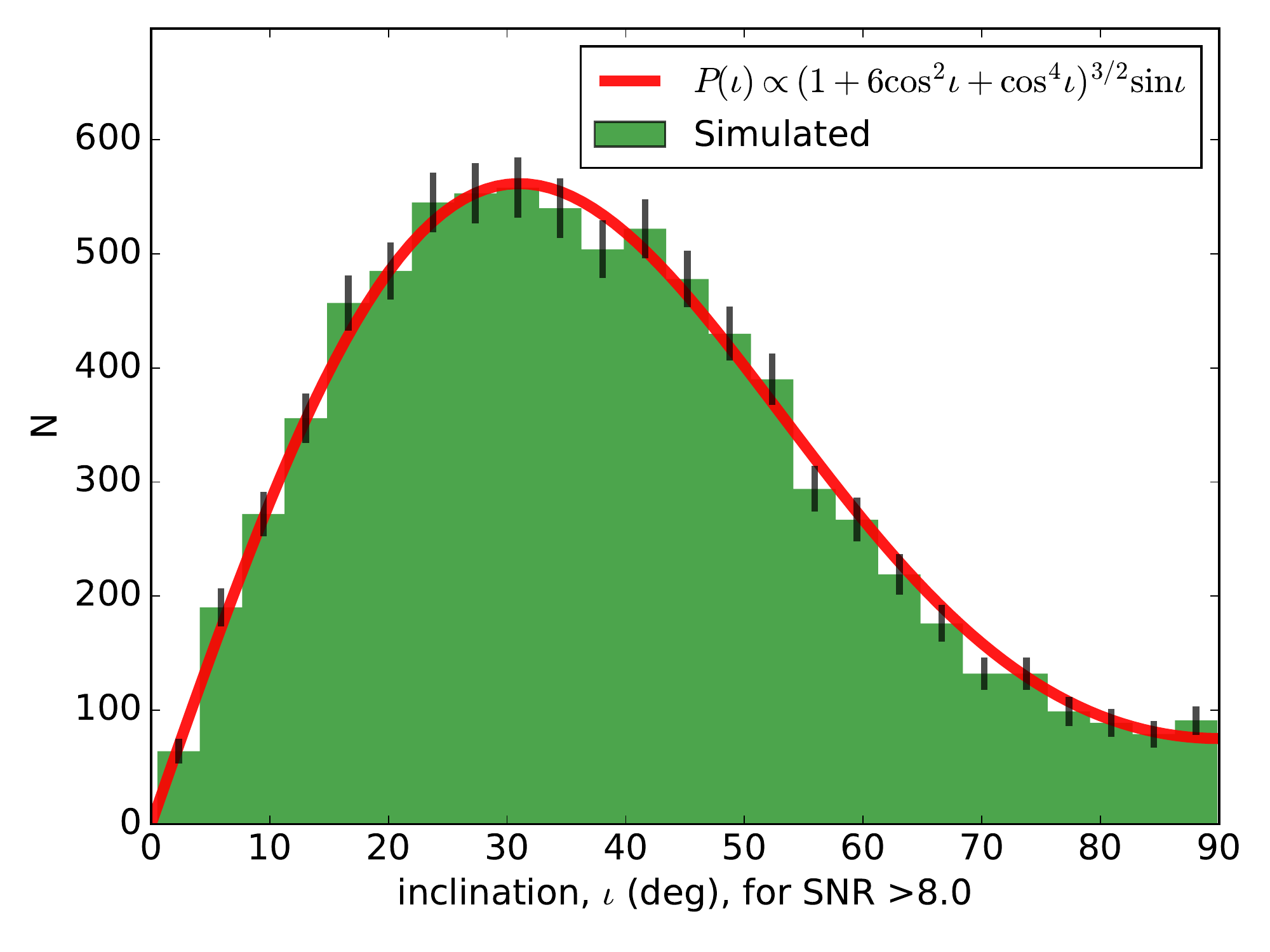}
\caption{Distribution of inclination angles from $\sim$ 8000 simulated BBHs
that had a SNR$>$8 out of $\sim$ 164,000 calculated waveforms. The initial
number of simulated BBHs was 4 $\times$ 10$^6$, most of which did not
contribute at SNR$>$8. The red curve shows the expected distribution. }
\label{fig:incl} \end{figure}

The probability that an observed GW from a compact binary
merger will have an inclination  $\iota$ is
\citep{Schutz11FoM}:
\begin{equation}\label{eq:incldistr}
P(\iota) \propto (1+6\cos^2 \iota + \cos^4 \iota)^{3/2} \sin \iota.
\end{equation}
We plot this distribution in Figure \ref{fig:incl} (red curve), overlaid with
the distribution of simulated GW signals as a function of
inclination angle (green histogram).  The figure  illustrates that our
simulations capture the expected inclination dependence of GW detections.

\subsection{GBM sensitivity}
The flux measured by {\it Fermi}-GBM for \gwgrb\, in the 10-1000 keV interval is
$F\approx 2.4\times 10^{-7} \erg \cm^{-2} \s^{-1}$.  The sensitivity of
{\it Fermi}-GBM varies as a function of the spacecraft's geographic location, the
source position in spacecraft coordinates \citep{Meegan+09gbm} and other
unmodeled factors. 

To estimate the sensitivity of GBM for a GRB with a particular flux, we use the
following method: We consider the fluxes of triggered GRBs in the {\it Fermi}-GBM
catalog \citep{Gruber+14cat,Bhat+16cat} and analyze the low-end tail of the
distribution. This means fluxes typically below $3\times 10^{-7}\erg
\cm^{-2} \s^{-1}$. GRBs below this flux level can potentially go undetected for GBM even if they occur within the field of view.
The lowest flux where a GRB was triggered is $\sim 7\times
10^{-8}\erg \cm^{-2} \s^{-1}$, while the flux above which essentially all GRBs
that are within GBMs field of view are detected is $\sim 3\times 10^{-7}\erg
\cm^{-2} \s^{-1}$.  In this flux interval we fit the distribution with a
one-sided Gaussian (i.e defined only below the peak) and find that such a
function gives a good description of the distribution of fluxes. Using this
function we employ an acceptance-rejection method to decide if a GRB with a
simulated flux is detected.  We illustrate this function as a gradient below
the histogram in Figures \ref{fig:nunu} through \ref{fig:egw}. Simulated flux values falling on darker hues
have a larger chance for acceptance. The blue dashed curve marks the 50\%
acceptance rate.

\begin{table*}
\begin{center}
\begin{tabular}{ccccccccc} 
 \hline
 BBH name 		& M$_1$ 	& M$_2$ 	& M$_f$ & $a_f$ & D$_L$/Mpc & $z$ & E$_{\rm GW}/M_\odot c^2$ &  SNR\\
 \hline
 GW150914 		& 35.6	& 30.6	& 63.1	& 0.69	& 430 	& 0.09&3.1	 & 23.7	\\
 GW151012 (1)	& 23.3	& 13.6	& 35.7	& 0.67	& 1060	& 0.21&1.5	 & 9.7 	\\
 GW151226 (2)	& 13.7	& 7.7	& 20.5 	& 0.74	& 440 	& 0.09&1.0	 & 13.0 	\\
 GW170104 (3)	& 31.0 	& 20.1 	& 49.1 	& 0.66	& 960 	& 0.19&2.2	 & 13 	\\
 GW170608 (4)	& 10.9	&7.6	& 17.8	& 0.69	& 320 	& 0.07&0.9  & 13	\\
 GW170729 (5)	& 50.6	&34.3	& 80.3	& 0.81	& 2750 	& 0.48&4.8  & 13	\\
 GW170809 (6)	& 35.2	&23.8	& 56.4	& 0.70	& 990 	& 0.20&2.7  & 13	\\
 GW170814 (7) 	& 30.7  &25.3  	& 53.4  & 0.72	& 580	& 0.12&2.7  & 15 	\\
 GW170818 (8) 	& 35.5  &26.8  	& 59.8  & 0.67	& 1020	& 0.20&2.7  & 15 	\\
 GW170823 (9) 	& 39.6  &29.4  	& 65.6  & 0.71	& 1850	& 0.34&3.3  & 15 	\\
 \hline
\end{tabular}
\caption{Central values for the observed BBH merger parameters, taken from
\citet{GWTC1}. Masses are in units of Solar masses. Numbers next to GW events
are for referencing them in figures.}
\end{center}
\label{tab:bbhlist}
\end{table*}

\subsection{Opening angles}
GRBs involve jetted emission \citep{Rhoads99jet,Harrison+99jet}. The energetics
of the GRBs strongly depend on the value of the opening angle.  Jet opening angles
are measured from the achromatic break in the afterglow lightcurve of GRBs. The
break occurs when the increasing angular size observed ($\approx 1/\Gamma$, where $\Gamma$ is the bulk Lorentz factor of the outflow) of
the emitting surface grows larger  the angular size ($\theta_j$).  This results
in a dearth of emitting surface and thus a steeper fading.  There is a better
handle on the  opening angles in long GRBs, but short GRBs have an increasing
number of measured jets \citep{Fong+15sGRB}. 

In our simulations, the  GRB that accompanies the GW signal is jetted with a
half opening angle $\theta_j$. If our line of sight to the jet axis
($\theta_{v}$) is less than $\theta_j$, we may in principle observe the GRB,
subject only to the instrument's sensitivity and observing conditions.

It is unclear how the opening angles of BBH counterparts may be distributed.
Here we discuss 3 cases: a fixed narrow 20$^\circ$ opening angle; a wide jet with $\theta_j=90 ^\circ$ opening angle corresponding to isotropic emission; a case where we associate
uniformly random opening angles between 10$^\circ$ and 40$^\circ$.  For
simplicity, we only consider top-hat jets, without any angular structure.  In
other words, we expect to detect GRBs for which the observer lies within the
jet and no flux is expected for observers outside of the jet.

For the observed jet opening angles, the uniform 10-40$^{\circ}$ distribution is the
most realistic. In addition, looking at the reported SNR in Table
\ref{tab:bbhlist}, the SNR$>$12 requirement is more in line with GW detections.
We thus consider this our benchmark case and highlight column 5 in Table
\ref{tab:ex1}.

\section{Possible paths to link GW and EM energy for binary black hole mergers}
\label{sec:em}

It is unclear if BBH mergers are accompanied by significant EM radiation, and
indeed the most likely scenario is that they occur in very low density
environments where it is difficult to extract e.g. the rotational energy of the
BHs and channel it to gamma-rays.  

We note that all the  outlined scenarios to extract gamma-rays from a BBH
merger suffer from non-trivial shortcomings, or critiques \citep{Lyutikov16gw}.
We consider the enumerated processes only as a guide to give an order of magnitude
answer to the fundamental question we are seeking to answer: after how
many BBH merger detections during O3, should we expect the next gamma-ray
counterpart, assuming a particular scenario for \gwgrb.

We use the {isotropic-equivalent} gamma-ray luminosity of GW150914-GBM, $L_{\gamma,{\rm iso}}=1.8\times 10^{49}
\erg \s^{-1}$ \citep{Connaughton+16gbmgw} as a scaling factor to guide our
models. 
Some scenarios involve the 
final spin and the radiated energy in form
of GWs ($E_{\rm GW}$).  
We use the formalism of \citet{Jimenez+17gwfit}
 to calculate $E_{\rm GW}$ and the final spin
 parameter $a_f$.  In this approach, these quantities are determined via
 polynomial functions based on the results of detailed general-relativistic
 simulations, e.g. \citet{Husa+16waveforms} \citep[see also ][ for an alternative approach]{Buonanno+08BHspin}.
For our simulations, we use the
formalism of \citet{Jimenez+17gwfit}.

{{As indicated before,} the gamma-ray luminosity of \gwgrb is an isotropic-equivalent luminosity. The physical models considered in this section however yield the total available luminosity. We account for the beaming correction between the two quantities by assuming the opening angle of \gwgrb is representative of the simulated BBH population.  Thus when normalizing any given model to the flux of \gwgrb, the correction between the isotropic-equivalent (measured) luminosity and the total (model) luminosity is included in the scaling factor (the gamma-ray production efficiency is accounted for in a similar way).  The fact that \gwgrb's opening angle may not be representative for the BBH population is one of the reasons we consider pessimistic and optimistic cases (Section \ref{sec:input}). This {is} a possible factor that can drive a dimmer or brighter expected population, given \gwgrb and a model.}

\subsection{Neutrino driven wind}
\label{sec:nunu}

Neutrino driven winds are routinely invoked as mechanisms launching the GRB jet
\citep[e.g.][]{Ruffert+98nu}. It implies a dense accretion disk that emits
neutrinos towards the axial region of the system where $\nu\bar\nu$ collisions
result in e$^\pm$ pairs that in turn drive a relativistic jet.  Due to the high
accretion rate required, \citet{Li+16gw} find that this mechanism is unlikely
to be at work in the case of \gwgrb. Nonetheless, since this is one of the
leading jet launching mechanisms, we consider the implication of this scenario
here. 

\begin{figure}
\centering
\includegraphics[width=0.99\columnwidth]{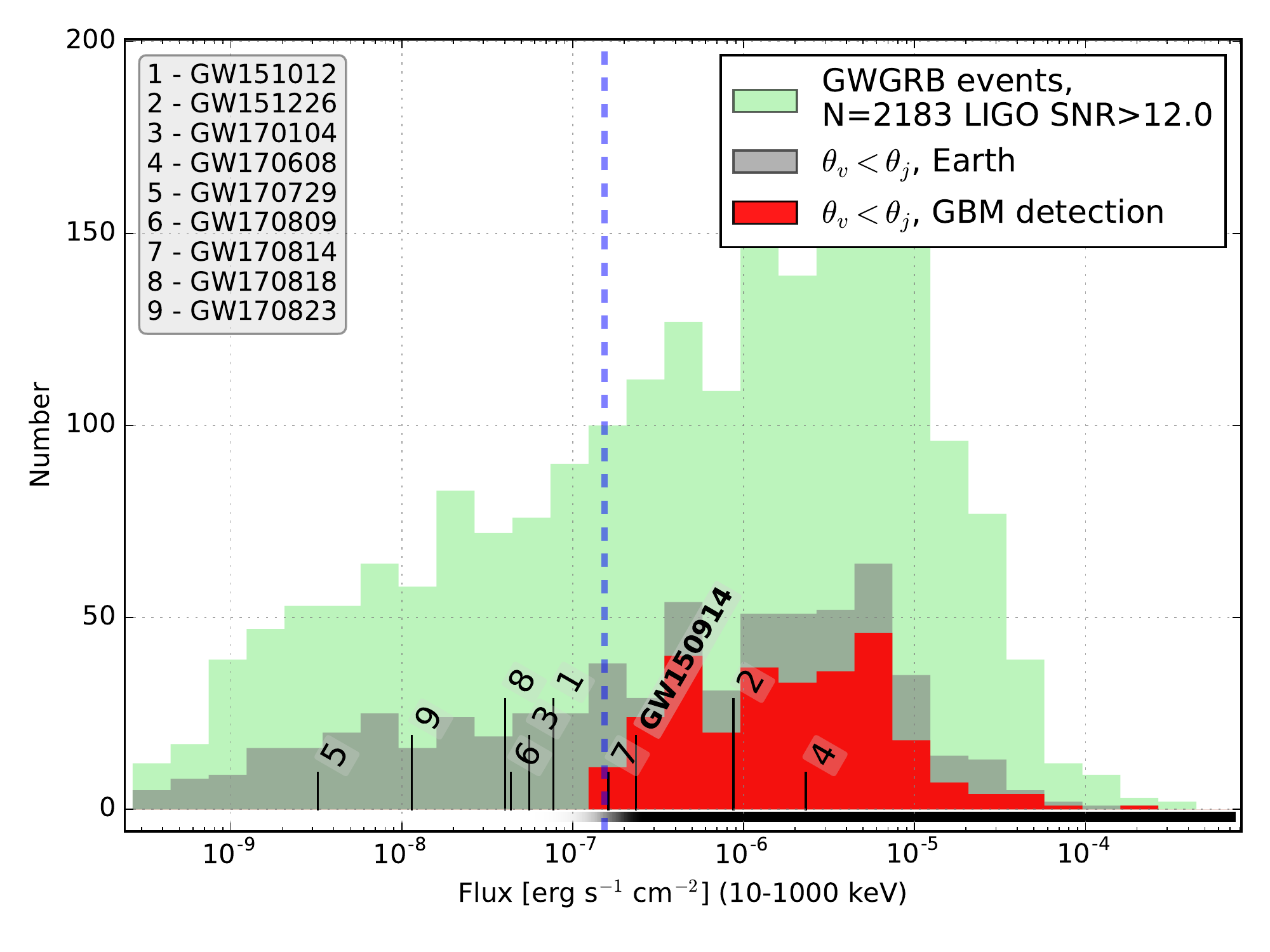}
\caption{Simulated EM counterparts based on neutrino driven wind model with GW
SNR$>$12 (see Section \ref{sec:nunu}). The model was normalized to exactly match \gwgrb's flux.  The green
histogram represents the simulated BBH mergers with LIGO-Virgo SNR$>$12.  The gray
histogram shows the cases when the associated jet points towards Earth. The red
histogram incorporates the sensitivity and the observing live-time of {\it Fermi}-GBM
showing all the detections from the simulations.  The band below the histogram
illustrates the GBM detection probability, and the vertical dashed blue line
indicates the 50\% mark. Based on this model every 7.7 BBH observed by LIGO-Virgo
should have a detectable EM counterpart by GBM. 4 out of the 10 observed BBH
mergers (GW170814 (7), GW150914, GW151226 (2) and GW170608 (4)) yield fluxes
above the GBM threshold (Section \ref{sec:obsBBH}).  } \label{fig:nunu}
\end{figure}

\citet{Zalamea+11nu} finds an empirical relation for the neutrino driven wind
luminosity:
\begin{equation}\label{eq:nunu}
L_{\nu\bar\nu}\approx 2.9\times 10^{47} \left(
\frac{f_3(a_f)}{f_3(0.68)}\right)^{-4.8}
\left(\frac{M_f}{62\,M_\odot}\right)^{-3/2} \left(\frac{\dot{M}}{M_{\odot}
\s^{-1}}\right)^{9/4} \erg \s^{-1}
\end{equation}
$f_3(a)=3+Z_2-[(3-Z_1)(3+Z_1+2Z_2)]^{1/2}$ and $Z_1=1+(2-a^2)^{1/3}[(1+a)^{1/3}
+ (1-a)^{1/3}] $ and $ Z_2=(3a^2+Z_1^2)^{1/2}$.  In this scenario, we vary the
uncertain accretion rate $\dot{M}$ and scale it to match the luminosity of
\gwgrb (additional details in Section \ref{sec:input}).
The results of simulations using this model are illustrated in Figure \ref{fig:nunu}

\subsection{Blandford Znajek (BZ) mechanism}
\label{sec:BZ}
\begin{figure}
\centering
\includegraphics[width=0.99\columnwidth]{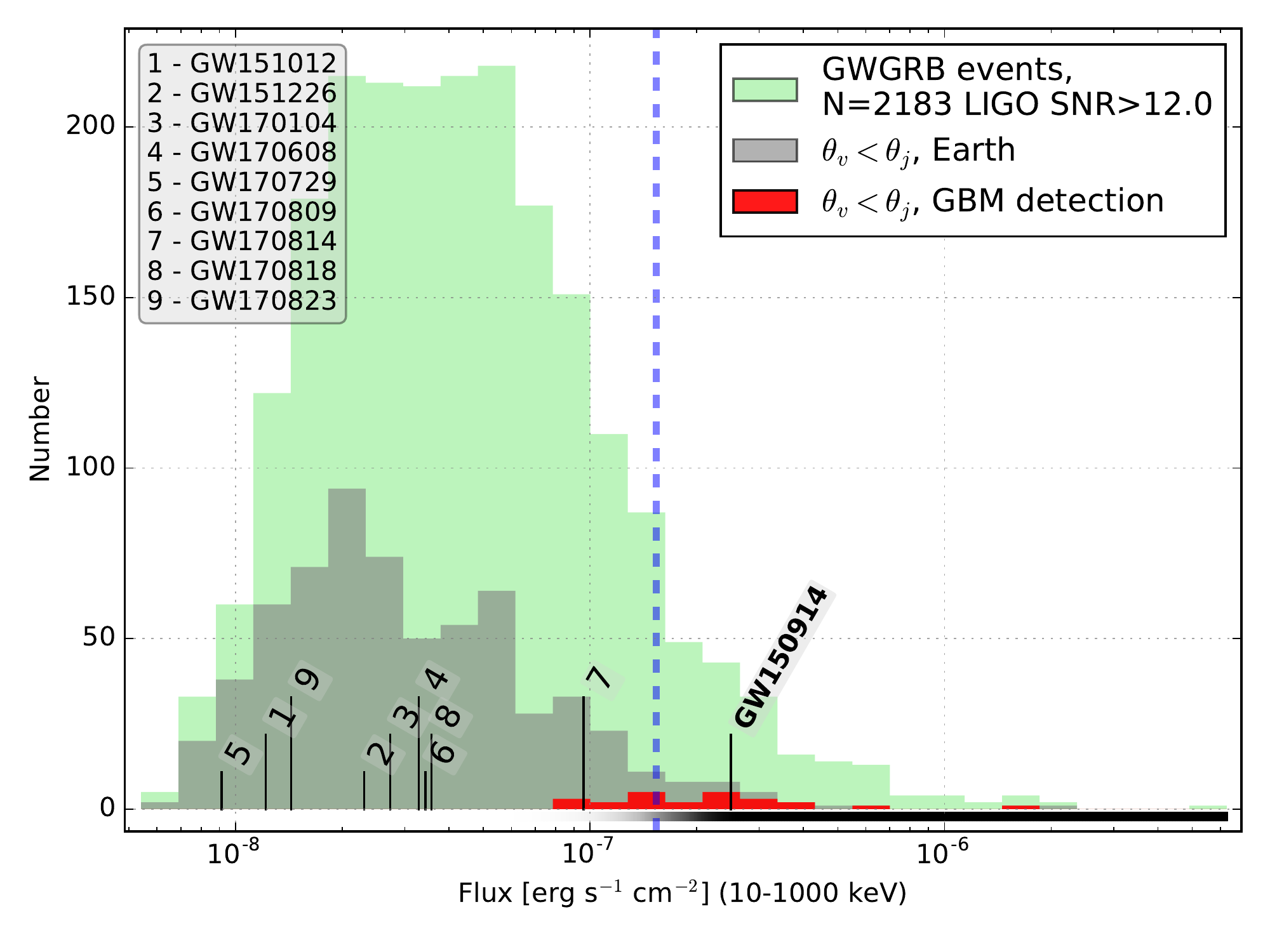}
\caption{Expected gamma-ray flux distribution from BBH, arising from
Blandford-Znajek mechanism (Section \ref{sec:BZ}). The BBH-to-GRB ratio is
2183/24 = 91.0. Notations are similar to Figure \ref{fig:nunu}.}
\label{fig:BZ}
\end{figure}

The BZ mechanism \citep{Blandford+77znajek} extracts the rotational energy ($
E_{\rm rot}=M_f c^2 \{1- \left[(1+(1-a_f^2)^{1/2})/2\right]^{1/2}\}$) of the BH
with the help of magnetic fields, where $a_f$ is the spin parameter of the
merged BH. Here we use the formulae by \citet{Komissarov+10BZ,Reynolds+06BZ}
for the extracted EM luminosity: 

\begin{equation}
L_{\rm BZ} = \frac{1}{3c} \left(\frac{\Psi_h \Omega_h}{4\pi}\right)^2 =
\frac{1}{12 c^3} G^2 M_f^2 B^2 f(a_f),
\end{equation}
where $\Omega_h\approx c^3/GM_f$ is the angular velocity of the BH, and $\Psi_h
\approx 2 \pi R_g^2 B$  the magnetic flux threading the  BH horizon at one of
its hemispheres, $R_g=GM_f/c^2$ is the gravitational radius, B the magnetic
field strength and $f(a)=f_1^4(a) f_2^2(a)$ ($f_1(a)= 2-a+2(1-a)^{1/2}$ and
$f_2(a)= a / 2 (1+(1-a^2)^{1/2})$) accounts for the approximations in the
expressions of $\Psi_h$ and $\Omega_h$.
The main unknown here is the  value of the magnetic field. We set this value so
that the simulations scale with the \gwgrb\, observations (see Section
\ref{sec:input} for a description of the scalings).  Figure \ref{fig:BZ} shows
the flux distribution obtained from this model.

\subsection{Charged black holes}
\label{sec:qq}

\citet{Zhang16chargedBH} proposed a mechanism, where at least one of the BHs
carries a significant charge and Poynting flux is extracted from the system {\citep[see also][for a similar approach]{2016ApJ...826...82L}}. To
approximate the luminosity in this scenario, we consider equation (7) of
\citet{Zhang16chargedBH}:
\begin{equation}
L_Q=C \frac{c^5}{G}\hat{q}^2 \hat{a}^{-15},
\end{equation}
where C=49/120000 a numerical factor, $\hat q$ is a dimensionless charge, in
units of critical charge, $Q_c$ ($Q_c=2\sqrt{G} M_f$) and $\hat a$ is the
normalized Newtonian distance of the two BHs, defined as $\hat a=(1+\sqrt{1-a^2})/2$. We
scale the luminosity to \gwgrb and vary the dimensionless charge to obtain the
gamma-ray flux of the simulated population.

\begin{figure}
\centering
\includegraphics[width=0.99\columnwidth]{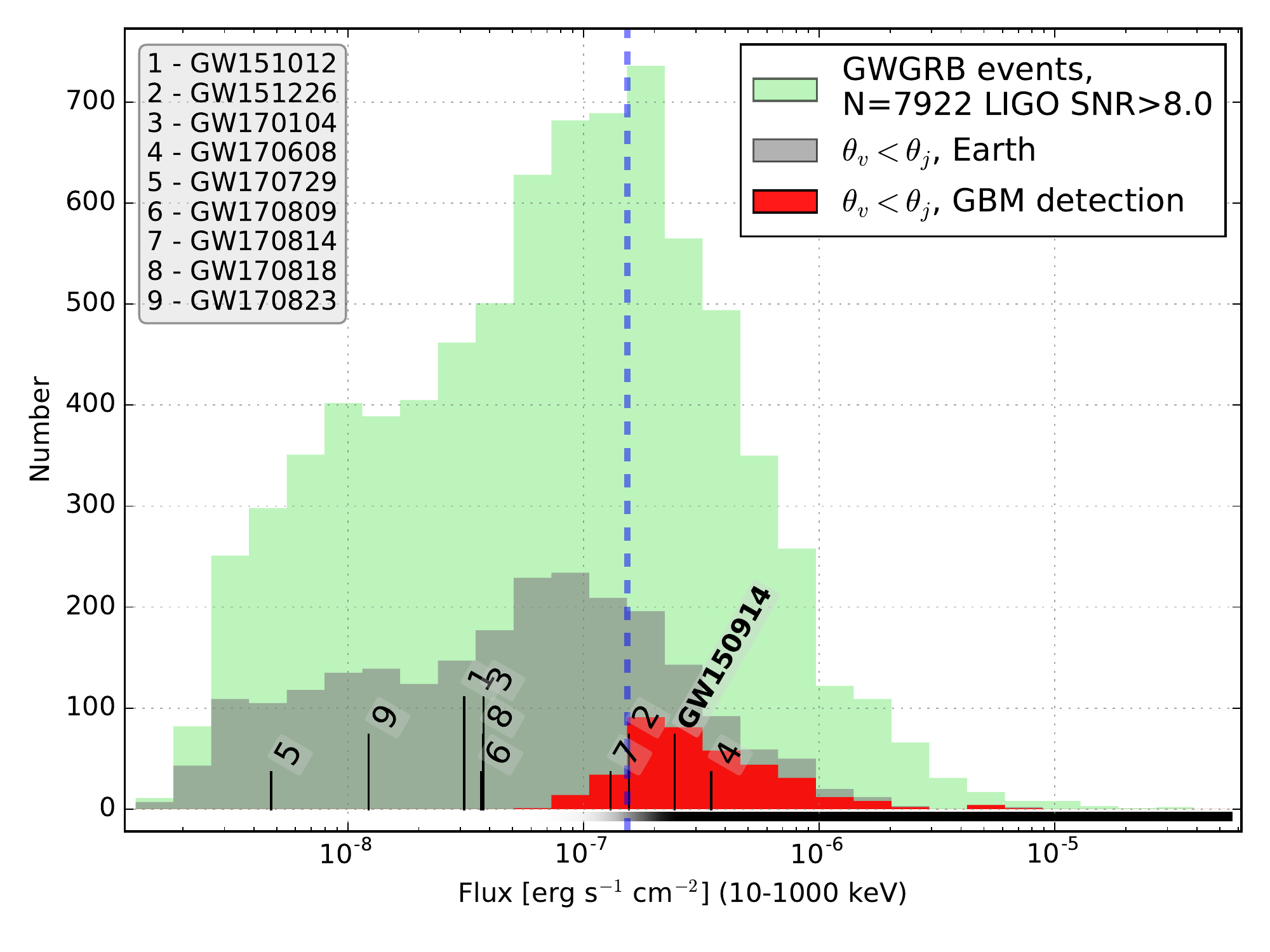}
\caption{Expected gamma-ray flux distribution from BBH, arising in the charged
BH scenario (Section~\ref{sec:qq}). Note that in this case we present GW signals with SNR>8. The
BBH-to-GRB ratio (ratio of numbers in the  green to red
histogram) is 7922/381 = 20.8. This is the expected number of BBH mergers GBM
has to follow until the detection of the next gamma-ray counterpart. Notations
are similar to Figure \ref{fig:nunu}.}
\label{fig:qq}
\end{figure}

\subsection{Gamma-rays as a fraction of GW energy release}
\label{sec:egw}
\begin{figure}
\centering
\includegraphics[width=0.99\columnwidth]{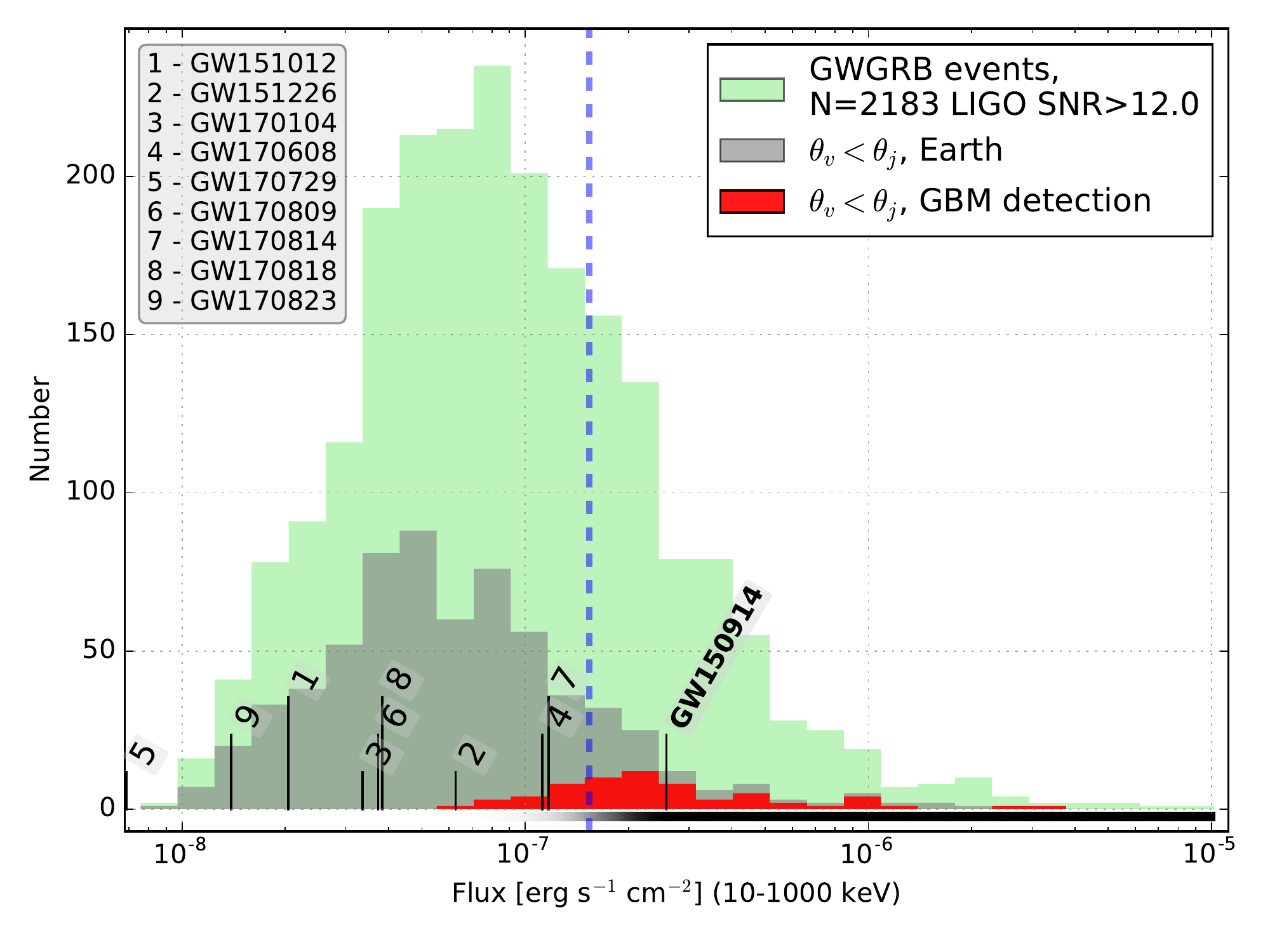}
\caption{Example of observed sources, where the EM energy is a fraction
$\epsilon_{\rm GW}=3.4\times10^{-6} \s^{-1}$ of the emitted GW radiation
($L_\gamma=\epsilon_{\rm GW} E_{\rm GW}$, Section \ref{sec:egw}). In
the above example the BBH-to-GRB ratio  is 34.1. Notations are similar to Figure \ref{fig:nunu}}
\label{fig:egw} \end{figure}

During the BBH merger a fraction of the mass is radiated in form of GWs. While
the coupling between GW and matter is extremely weak, here we explore a
scenario where the gamma-ray output of the system depends on the GW energy
output. We do not imply that the GW energy is channeled into gamma-rays, but
the gamma-ray generation correlates with the emitted GW energy.
We can parametrize the gamma-ray luminosity as a fraction of $E_{\rm GW}$,
$L_\gamma=\epsilon_{\rm GW} E_{\rm GW}$. \citet{Abbott+16gw150914properties}
estimates the total energy emitted in GWs: $E_{\rm GW}=5.3\times 10^{54} \erg$.
To normalize the GW energy to e.g. \gwgrb, we have $\epsilon_{\rm GW} =
3.4\times 10^{-6} \s^{-1}$. We refer to this model as 'gamma-GW fraction' and
show the associated gamma-ray flux distribution in Figure \ref{fig:egw}.

\subsection{Power law dependence on the final mass}
\label{sec:PL}
For this scenario, we assume the EM luminosity, $L_\gamma$ has a power law
dependence on the final BH mass, $M_f$: $L_\gamma =L_* ( M_f/M_*)^{p}$. We may
use e.g.  $L_*=1.8\times 10^{49} \erg \s^{-1}$ and $M_*=62 M_\odot$ for
normalization.

\begin{figure}
\centering
\includegraphics[width=0.99\columnwidth]{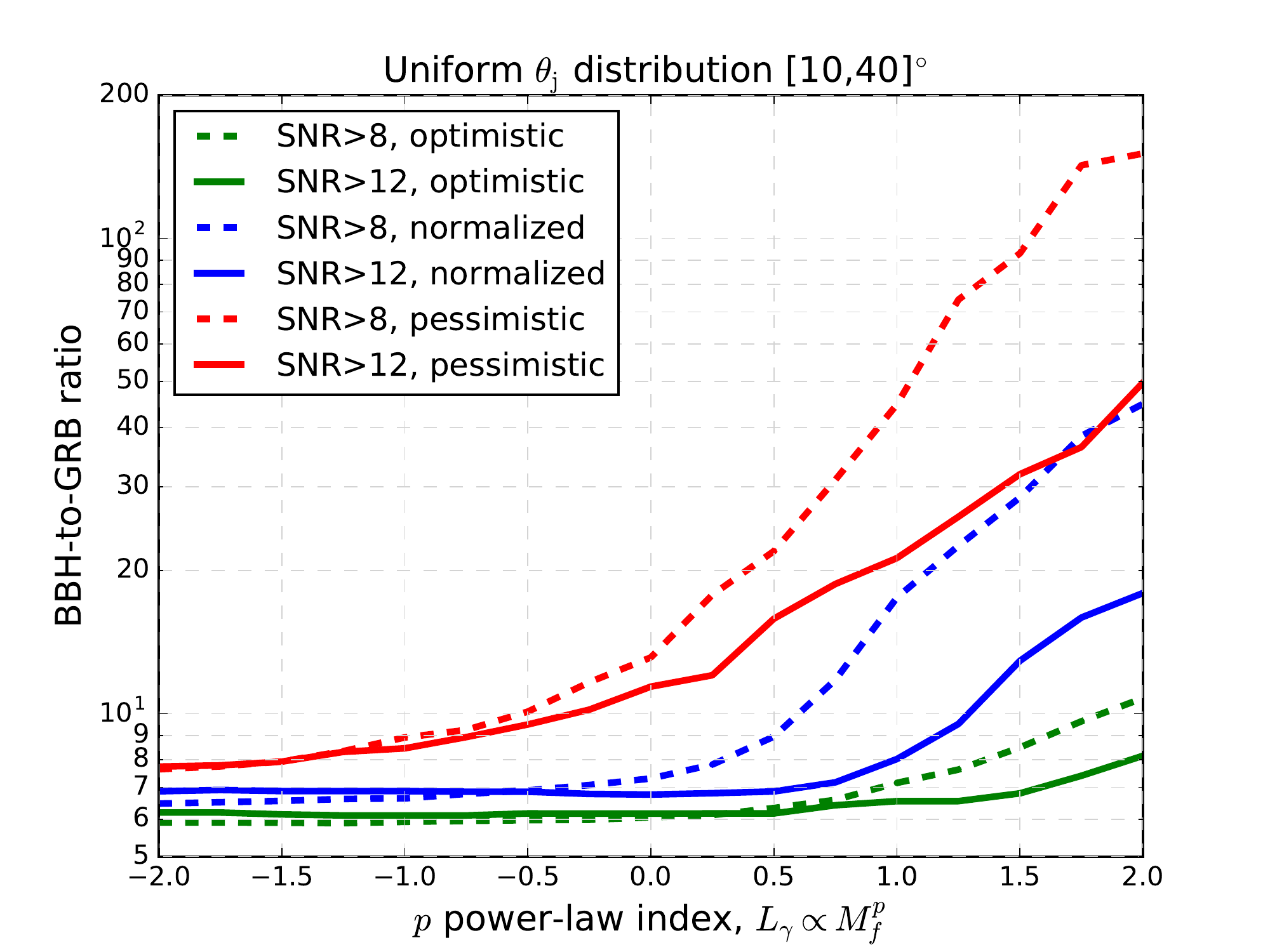}
\caption{Effect of changing the power law index on the BBH-to-GRB ratio. We
assume the opening angles have a uniform distribution between 10 and 40
degrees. The lines are averages of 4 random  realizations of BBH orientations.}
\label{fig:PL}
\end{figure}

This scenario is best suited to illustrate the scaling with final mass for the
gamma-ray flux (see Figure \ref{fig:PL}). The final mass of \gw was at the high
end of the detected mass distribution and  the source was relatively nearby. If
$L_\gamma$ correlates positively with $M_f$ or $p>0$, we expect gamma-ray
counterparts to be a rare occurrence. If however the gamma-ray emission is
negatively correlated with the final BH mass ($p<0$), that would suggest that
\gwgrb\, was a common event. 

\subsection{Input for simulations}
\label{sec:input}
For each emission model described above, we consider the following subcases:
\begin{enumerate}
\item {\it Pessimistic: } In this case we simulate the unmodeled parameters
(magnetic field, $B$ for BZ; accretion rate, $\dot{M}$ for neutrino
annihilation; normalized charge $\hat q$ for the charged BHs, etc) for each
scenario according to a log-normal distribution with a mean that is half order
of magnitude ($10^{1/2}$) lower than what is required to produce a \gwgrb-like
signal, and has a standard deviation of half an  order of magnitude
($10^{1/2}$).

\item {\it Normalized to \gwgrb:} The parameters of the scenario are chosen so
that they reproduce {\it on average}, the {\it Fermi}-GBM luminosity of \gwgrb,  with
a standard deviation of half an order of magnitude.

\item {\it Optimistic: } This  is similar to the pessimistic case, but the mean
of the log-normal distribution is half an order of magnitude {\it higher} than
the one needed to reproduce \gwgrb.

\item In addition to the above three cases, we also carry out simulations where the
gamma-ray luminosity is required to match {\it exactly}  \gwgrb. The purpose
such a scenario is to gauge if currently observed  BBHs would be expected to
produce a detectable gamma-ray signal (for the \bbgr ratio in these cases see
Figures \ref{fig:nunu},
through \ref{fig:qq}, and Section \ref{sec:current}).

\end{enumerate}

\subsection{Gamma-rays from a random,  unknown process}
\label{sec:rnd}
It is possible that a hereto unknown process is responsible for \gwgrb, that is
unrelated to the GW observables {\citep[see e.g.][ for BBH mergers occurring in active galactic nuclei]{2017ApJ...835..165B}}.
We explore this scenario by randomly assigning
gamma-ray luminosity to simulated BBH mergers.
We constrain the mean and standard deviation of the flux to the 3 scenarios outlined in Section \ref{sec:input} (normalized, pessimistic and optimistic), but independent of any physical property of the systems.

\begin{table*}
\begin{center}
\caption{The  BBH-to-GRB  ratio in different scenarios. We consider three types
of opening angles: fixed narrow ($\theta_j=20^\circ$), distributed randomly
between 10 and 40 degrees and isotropic ($\theta_j=90^\circ$). We show two
limits in GW SNR (8 and 12). The energy extraction scenarios are described in
Section \ref{sec:em}, while the 3 cases are presented in Section
\ref{sec:input}. Column 5 is highlights the benchmark case, indicating the most
similarity to short GRB opening angles and GW observations.}
\begin{tabular}{|l|c||r|l||r|>{\bfseries}l||r|l|} 
\hline
Scenario & Section	& \multicolumn{2}{c||}{$\theta_j=20^\circ$ }   &  \multicolumn{2}{c||}{$\theta_j=$Unif.$[10,40]^\circ$}     &  \multicolumn{2}{c|}{$\theta_j=90^\circ$}  \\\cline{3-8}
    &	& SNR$>$8 & SNR$>$12  & SNR$>$8 & SNR$>$12  & SNR$>$8 & SNR$>$12  \\
		\hline
neutrino wind, pessimistic & & 31.1 & 29.1 & 20.7 & 15.3 & 5.6 & 3.9 \\
neutrino wind, normalized  &\ref{sec:nunu} & 15.7 & 14.6 & 10.4 & 8.6 & 2.8 & 2.5 \\
neutrino wind, optimistic & & 10.7 & 10.5 & 7.1 & 6.5 & 2.0 & 1.9 \\
		\hline
Blandford Znajek, pessimistic & & 304.7 & 181.9 & 146.7 & 94.9 & 33.4 & 24.0 \\
Blandford Znajek, normalized  &\ref{sec:BZ} & 44.5 & 32.6 & 34.9 & 21.8 & 7.1 & 5.0 \\
Blandford Znajek, optimistic	& & 16.4 & 14.1 & 10.8 & 9.0 & 2.9 & 2.4 \\
\hline
Charged BH, pessimistic & & 158.4 & 121.3 & 101.6 & 66.2 & 22.3 & 13.9 \\
Charged BH, normalized  & \ref{sec:qq}& 22.6 & 21.0 & 14.6 & 11.7 & 3.9 & 2.9 \\
Charged BH, optimistic  & & 10.0 & 9.7 & 6.5 & 5.9 & 1.9 & 1.8 \\
		\hline
$L_{\gamma}$=E$_{\rm GW} \times {\epsilon_{\rm GW}}$, pessimistic & & 720.2 & 218.3 & 565.9 & 121.3 & 55.4 & 21.6 \\
$L_{\gamma}$=E$_{\rm GW} \times {\epsilon_{\rm GW}}$, normalized &\ref{sec:egw} & 176.0 & 62.4 & 99.0 & 29.1 & 11.7 & 5.1 \\
$L_{\gamma}$=E$_{\rm GW} \times {\epsilon_{\rm GW}}$, optimistic & & 31.7 & 17.1 & 18.3 & 9.2 & 3.7 & 2.2 \\
		\hline
${L_{\gamma}} \propto {\rm M}_{\rm f}^{-2.0}$, pessimistic  & & 16.9 & 15.0 & 11.0 & 8.7 & 3.0 & 2.5 \\
${L_{\gamma}} \propto {\rm M}_{\rm f}^{-2.0}$, normalized &\ref{sec:PL}	& 13.4 & 12.3 & 9.1 & 7.4 & 2.5 & 2.2 \\
${L_{\gamma}} \propto {\rm M}_{\rm f}^{-2.0}$, optimistic &	& 11.9 & 10.9 & 7.8 & 6.9 & 2.2 & 2.0 \\
		\hline
${L_{\gamma}}$=$5.7\times10^{48}$ erg s$^{-1}$(const.) & & 114.8 & 54.6 & 64.9 & 28.7 & 10.7 & 5.0 \\
${L_{\gamma}}$=$1.8\times10^{49}$ erg s$^{-1}$(const.) &\ref{sec:PL} & 31.1 & 18.3 & 19.6 & 10.2 & 4.2 & 2.7 \\
${L_{\gamma}}$=$5.7\times10^{49}$ erg s$^{-1}$(const.) & & 14.6 & 11.8 & 9.6 & 7.2 & 2.5 & 2.0 \\
   	\hline
$L_{\gamma}\propto$  random, pessimistic& & 79.2 & 42.8 & 49.2 & 22.1 & 8.8 & 4.8 \\
$L_{\gamma}\propto$  random, normalized & \ref{sec:rnd} & 29.8 & 20.6 & 18.0 & 11.2 & 4.1 & 2.8 \\
$L_{\gamma}\propto$  random, optimistic& & 15.4 & 12.5 & 9.6 & 7.6 & 2.6 & 2.1 \\
		\hline
\end{tabular}
\end{center}
\label{tab:ex1}
\end{table*}
\bigskip

\subsection{Application to currently observed BBH mergers and GBM upper limits}
\label{sec:current}

There are 10 BBH mergers detected prior to the start of O3 \citep{GWTC1}, with
only one putative gamma-ray counterpart. We note that in this work we used the
O3 sensitivity for the simulations that is superior to the previous two
observing run's sensitivity.  The \bbgr is expected to be lower for BBHs
detected in a less sensitive configuration, because of the lower average
distance of the sources imply higher gamma-ray flux. At the same time the expected number
of events is also lower.

Gamma-ray emission will likely be beamed, while the GW emission is close to isotropic. Thus for individual events it is difficult to rule out any particular model. Even if a model suggests a large flux, it is possible that the jet was beamed away from Earth, thus undetectable for GBM. 
This could be the case e.g. for GW170814 (7),  GW151226 (2) and GW170608 (4) in Figure \ref{fig:nunu}. The neutrino driven wind model predicts a detectable flux for these events. Interestingly these three BBH mergers imply fluxes within the sensitivity of GBM for the gamma-GW fraction model (Figure \ref{fig:egw}), and the charged BH scenario (Figure \ref{fig:qq}) as well. For the BZ model, only  GW170814 (7) produces detectable gamma-ray flux (Figure \ref{fig:BZ}). 

Fermi GBM reported on a weak source following GW170814 (7) \citep{Goldstein+17gcn}, but it was deemed unassociated with GW170814, based on the small overlap between source locations. Part of the location region of GW151226 (2) was occulted for Fermi and no significant sources were detected \citep{Burns+15gcn}. The location region of GW170608 (4) was covered well by GBM and no sources were reported \citep{Hamburg+17gcn}.

\section{Discussion}
\label{sec:disc}
We have calculated the expected LIGO-Virgo detected BBH mergers to gamma-ray
counterparts (\bbgr) that could be observed by {\it Fermi}-GBM in different
scenarios.  The results of this study are summarized in Table \ref{tab:ex1}.

The magnitude of the gamma-ray flux depends on the adopted scenario.
Among the BBHs detected by LIGO and Virgo \gw has a relatively large final mass
and it is located close compared to other GW events. It is thus in line with
expectations, that in scenarios where the gamma-ray luminosity positively
correlates with the mass of the final BH we expect a large number of BBH
observations before another counterpart is detected. Conversely, for the
scenarios where the EM luminosity is inversely proportional with the final
mass, the required number of BBH observations before another gamma-ray
counterpart is observed is low ($\lesssim$ 10). This is displayed in Figure
\ref{fig:PL}: for negative $p$ values, essentially all the generated GRB flux
that reaches the observer is detected. The \bbgr is governed by the
opening angle distribution. As we move to positive $p$ values, the ratio
increases because we scale our calculations to \gw, involving a nearby, massive
BH. For $p>0$ the bulk of the simulated BHs will be less massive thus generating
lower gamma-ray flux.

Figures \ref{fig:nunu} through \ref{fig:egw} illustrate the calculation of the
\bbgr in the cases where the assumed mechanism reproduces \gwgrb exactly. The
green histogram shows the gamma-ray flux for all the detectable BBHs with SNR
12 or 8 for LIGO-Virgo. The initial cut for the jets that point elsewhere
reduce the numbers to the gray histogram. Considering the {\it Fermi}-GBM live-time, sky
exposure and detection threshold further reduces the detected GRBs to the red
histogram. For each scenario, we show the expected fluxes from the observed BBH
population as well, indicated by numbers on the histograms.

\subsection{Favored and unfavored mechanisms}
We find that the {\it neutrino driven wind} mechanism gives the lowest \bbgr
(comparable to the power low scenario with $p=-2$). This can be understood as
we scale the gamma-ray flux to the relatively nearby and high mass \gw event.
In this scenario there is a strong dependence on the final mass (Equation
\ref{eq:nunu}).  Even in the pessimistic scenario we can constrain this
mechanism (rule it out) after $\sim$15 BBH observations.

If the EM output depends on the final BH mass as a power law, our
simulations show that for positive power law indices ($p$, where
$L_\gamma\propto M_f^p$) the \bbgr ranges from $\sim$ 5 to a few times 10 (for
$p=2$ and SNR$>$12) or to $\lesssim$ 200 (for $p=2$ and SNR$>$8) as shown in
Figure \ref{fig:PL}.

\subsection{Observed BBHs}
\label{sec:obsBBH}
The neutrino driven wind scenario results in  3 (out of 9) of the observed GWs
with confidently detectable flux for GBM (Figure \ref{fig:nunu}) while the
simulations give a \bbgr of 7.7. The charged BH scenario gives similarly 3 GWs
in the detectable flux range for GBM, albeit with lower flux (see Figure
\ref{fig:qq}). The gamma-GW fraction scenario places 3 GW in the marginally
detectable flux regime, but well below the 50 \% detection threshold (see
Figure \ref{fig:egw}).  The BZ scenario results in detectable counterpart for a
single known GW (see Figure \ref{fig:BZ}).  Taken at face value these results
indicate that with an increasing number of non-detections, the neutrino driven
wind and the charged BH model can be ruled out first, then the gamma-GW
fraction scenario. Finally the BZ mechanism requires the most non-detections
for it to be ruled out.

\subsection{Role of the opening angle and signal to noise limit}
The assumed opening angle significantly change the resulting \bbgr: wider jets
are more likely to include the detector in their aperture thus decreasing the
\bbgr. We focused the discussion so far on the benchmark case of uniformly
distributed opening angles between 10 and 40$^{\circ}$. We have also calculated
a narrow (20 degree) and wide, 90 degrees case as well, to get a sense of how the
number of GRBs will change.

The narrow opening angle increases \bbgr compared to the benchmark case by
anywhere between 50\% to 100\%. The wide, isotropic emission case decreases
\bbgr by a factor of 3 and up to 20.

Allowing for lower SNR (>8) threshold has the effect of increasing the number
of GWs detected and the distance limit. This in turn disproportionately
increases the number of gamma-ray events with flux below the detection
threshold, thus increasing the \bbgr. The increase in \bbgr can range from a
few percent to a factor of 3 (see Table \ref{tab:ex1}).

\subsection{Triggers versus off-line searches}
It is important to note that the GBM sensitivity considered here applies to
GRBs that were triggered real-time on-board the spacecraft. \gwgrb was found in
an off-line search.  Off-line searches improve the sensitivity by a factor of
few \citep{Kocevski+18targeted}, thus improving (decreasing) the \bbgr.

\subsection{Expected distribution of \bbgr}

A model predicts on average $N_{\rm BGR}$ BBH observations for each GRB. The probability of observing $0$ GRBs after $N_{\rm BBH}$ BBH detections for such a model, can be calculated from the binomial distribution:
\begin{equation}
\label{eq:prob}
P\left(x=0\middle|N_{\rm BBH},p=\frac{1}{N_{\rm BGR}}\right)=\left(1-\frac{1}{N_{\rm BGR}}\right)^{N_{\rm BBH}}.
\end{equation}

We illustrate the implications for the models by calculating this probability after {\it exactly} the same number of observed BBHs as the model's \bbgr would suggest ($N_{\rm BBH}=N_{\rm BGR}$). Using representative cases, e.g. $N_{\rm BGR} = \{5,10,20, 100\}$, yields the following probabilities: $
P(x=0|N_{\rm BBH}=N_{\rm BGR}) = \{ 0.33\,   ,  0.35, \, 0.36,\,  0.37\}$. In other words, after 
$N_{\rm BGR}$ non-detections there is still a $\sim 35$\% probability that the model is correct and will lead to a BBH observation with a counterpart.

After  $N_{\rm BBH}>N_{\rm BGR}$ observations with no GRB counterparts, what is the probability that the model predicting $N_{\rm BGR}$ BBHs for each GRB on average is correct? 
Equation \ref{eq:prob}  can be easily inverted to calculate the number of BBH observations necessary to rule out a model predicting $N_{\rm BGR}$, with a desired confidence $P_{\rm conf}$:
\begin{equation}
N_{\rm BBH}=\frac{\log P_{\rm conf} }{ \log \left(1-\frac{1}{N_{\rm BGR}}\right) } \approx -N_{\rm BGR} \log P_{\rm conf}.
\end{equation}

\begin{figure}
\centering
\includegraphics[width=0.99\columnwidth]{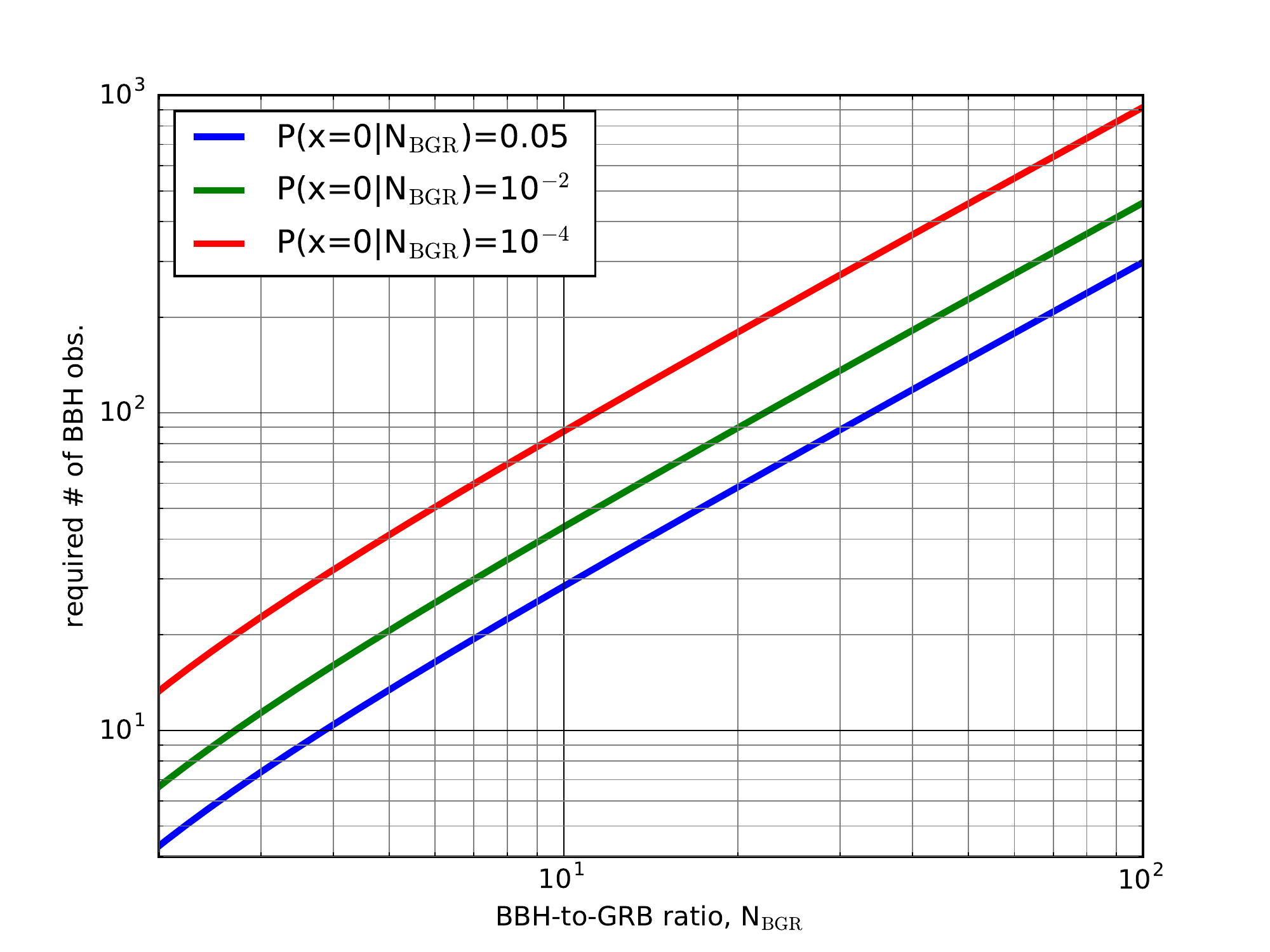}
\caption{Number of BBH observations with no GRB counterparts required to rule out a model that predicts a given \bbgr. The different curves correspond to different confidence levels.}
\label{fig:prob}
\end{figure}

For example, a model predicting \bbgr of $N_{\rm BGR}=10$ can be ruled out with $P_{\rm conf}=10^{-2}$ confidence after $\approx 44$ non-detections. Figure \ref{fig:prob} shows the required number of BBH observations to rule out a model with an arbitrary \bbgr of $N_{\rm BGR}$ with a given confidence.

\subsection{Detection prospects}
Assuming a rate of 40 BBH detections during the year long observing run O3
\citep{GWTC1,Abbott+16GWsens}, and with no counterpart detection, we can make
the following statements for the benchmark case ($\theta_j$ uniformly
distributed random between 10-40$^\circ$ and SNR>12, highlighted in Table
\ref{tab:ex1}): The neutrino driven wind, the power law model with $p=-2, 0$,
and the random model can be ruled out even in the pessimistic case. The BZ
model, the charged BH, the gamma-GW fraction models can be constrained in all
but the pessimistic cases. 

Assuming again the benchmark case, the majority of
the scenarios point to 6 to 20 BBHs for every gamma-ray counterpart detected.
This is encouraging for the third observing run of LIGO-Virgo, that is planned
to last a year, and is predicted to observe about 40 BBH mergers. If no
counterpart is detected during O3, most of the benchmark models can be ruled
out.  The only feasible models remaining after O3 are those that assume that
\gwgrb was a brighter than usual event given a particular gamma-ray producing
mechanism (pessimistic variants of e.g. Blandford-Znajek, charged BH or
gamma-GW fraction models).

After 40 BBH observations with no counterparts, almost all isotropic emission
models ($\theta_j=90^\circ$) can be ruled out with reasonable confidence.  For
narrowly beamed gamma-ray emission ($\theta_j=20^\circ$), most models can be
ruled out, but some can only be constrained.

\section{Conclusion}
\label{sec:concl} 
In this paper we investigated detection prospects of gamma-ray emission
associated with BBH mergers, using different emission models.  We emphasise
here the caveat that EM counterparts from stellar mass binary BH mergers are
unexpected and all the outlined models have considerable issues to be worked
out. Nonetheless, the detection of \gwgrb presents and intriguing prospect that
through some mechanism, not considered before, significant energy might be
channeled from GW to gamma-rays in the process of a BBH merger.

Assuming GW150914-GBM was a GRB associated with GW150914, we quantitatively estimated how many BBH events {\it Fermi}-GBM should follow up in order to detect a second one with a counterpart. 
We find that for a majority of the models we expect 6-20 BBH mergers for every
gamma-ray counterpart. Hence we expect that, after the ongoing third LIGO-Virgo
observing run, either another counterpart will be found, or many of the models
discussed here will be incompatible with the data.

{\bf Acknowledgements: } We thank Bing Zhang for valuable input, Francesco Pannarale for comments on the manuscript {and the anonymous referee for a valuable comment}. PV acknowledges  support from Fermi grant NNM11AA01A and 80NSSC17K0750. TD and EB are supported by an appointment to the NASA Postdoctoral Program at the Goddard Space Flight Center, administered by Universities Space Research Association under contract with NASA. NC is supported by NSF Grants PHY-1806990 and PHY-1505373.

This paper has LIGO document number LIGO-P1900147.

\bibliographystyle{yahapj}

\end{document}